\def\aeti{$\alpha$-(BE\-DT\--TTF)$_2$\-I$_{3}$}

\def\cm{cm$^{-1}$}
\documentclass[aps,prb,twocolumn,showpacs]{revtex4}
\usepackage{graphicx}%
\usepackage{graphicx}%
\usepackage{dcolumn}%
\usepackage{bm}%
\usepackage{xcolor}

\begin{document}

\title{Pressure-dependent optical investigations of $\alpha$-(BEDT-TTF)$_2$I$_3$: \\
tuning charge order and narrow gap towards a Dirac semimetal}
\author{R. Beyer}       %\author{Rebecca Beyer}
\author{A. Dengl}       %\author{Armin Dengl}
\author{T. Peterseim}   %\author{Tobias Peterseim}
\author{S. Wackerow}    %\author{Sabine Wackerow}
%\author{M. Herter}     %%\author{Melina Herter}
\author{T. Ivek}        %\author{Tomislav Ivek}
\author{A.V. Pronin}    %\author{Artem V. Pronin}
\author{D. Schweitzer}  %\author{Dieter Schweitzer}
\author{M. Dressel}     %\author{Martin Dressel}
\affiliation{Physikalisches Institut, Universit\"at Stuttgart, Pfaffenwaldring 57, 70550 Stuttgart, Germany}
\date{\today}

\begin{abstract}
Infrared optical investigations of $\alpha$-(BEDT-TTF)$_2$I$_3$ have been performed
in the spectral range from 80 to 8000~cm$^{-1}$ down to temperatures as
low as 10~K by applying hydrostatic pressure.
In the metallic state, $T > 135$~K, we observe a 50\% increase in the Drude contribution
as well as the mid-infrared band due to the growing intermolecular orbital overlap with pressure up to 11~kbar.
In the ordered state, $T<T_{\rm CO}$, we extract how the electronic charge per
molecule varies with temperature and pressure:
Transport and optical studies demonstrate that charge order and metal-insulator
transition coincide and consistently yield a linear decrease of the transition temperature $T_{\rm CO}$ by $8-9$~K/kbar. The charge disproportionation $\Delta\rho$ diminishes by
$0.017~e$/kbar and the optical gap $\Delta$ between the bands decreases with pressure by -47~cm$^{-1}$/kbar.
In our high-pressure and low-temperature experiments,
we do observe contributions from the massive charge carriers
as well as from massless Dirac electrons
to the low-frequency optical conductivity, however, without being able to disentangle them unambiguously.
\end{abstract}

\maketitle

\section{Introduction}
The two-dimensional organic conductor \aeti\
had been subject to intense studies for
quite some time\cite{Bender84a,Bender84b,Dressel94}
before the metal-to-insulator transition at $T_{\rm
CO}=135$~K was recognized as the entrance of a charge-ordered
state.\cite{Kino95,Kino96,Seo04,Takano01,Wojciechowski03,Dressel04}
Recently, doubt has been casted whether intersite electron-electron repulsion
actually is the driving force for the observed charge disproportionation
since it became clear that the anions play a major role and might trigger the rearrangement of charge on the donors.\cite{Alemany12}
The application of pressure gradually shifts
the metal-insulator transition and the system remains metallic
down to lowest temperatures.\cite{Kartsovnik85,Schwenk85,Tajima00}
The nature of this state as well as the detailed mechanism driving the transitions, however, are far from being fully understood.

Vibrational spectroscopy is probably the most suitable methods to determine the
charge per molecule quantitatively
and with high accuracy;\cite{Drichko09,Girlando11,Yakushi12}
our present pressure-dependent infrared experiments enable us to trace the
charge disproportionation of \aeti\ down to low temperatures
as the transition is suppressed. On the one hand
pressure enhances the coupling between the
organic BEDT-TTF molecules and the I$_3^-$ anion layer via the hydrogen bonds;
on the other hand it also modifies the orbital overlap
making electronic correlations less effective.
Thus we extract information on the bandwidth
from the spectral weight redistribution observed in the electronic
part of the optical spectra as a function of pressure and temperature.

\begin{figure}[tp]
    \centering
    \includegraphics[width=1\columnwidth]{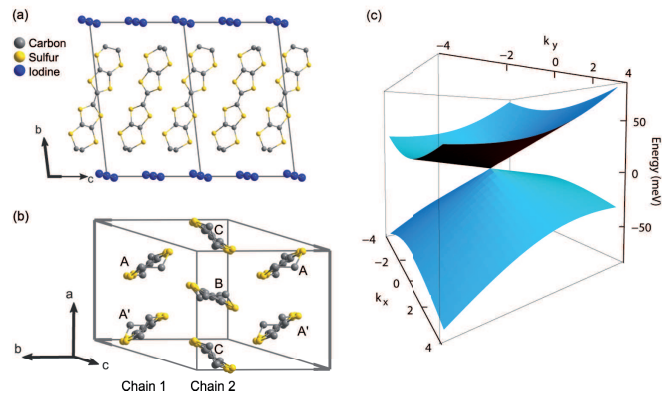}
    \caption{(Color online) The structure of
$\alpha$-(BEDT-TTF)$_{2}$I$_{3}$, where BEDT-TTF stands for
bis\-ethylene\-di\-thio-tetra\-thia\-fulvalene. Blue: I, gray: C,
yellow: S, for clearity, the hydrogen atoms have been omitted.
(a)~View on the planes of BEDT-TTF separated by I$_{3}^-$ sheets.
(b)~Out-of-plane view along the long axes of the molecules, which
form two chains: chain 1 consists of molecules A and A$^{\prime}$
and chain 2 includes molecules B and C.
%By stoichiometry the BEDT-TTF molecules are supposed be positively charge by $0.5e$, however, the later are carry slightly different charges even at high temperatures.
(c)~Sketch of the band dispersion at high-pressure in the first Brillouin zone around one contact point  (after Ref.~\onlinecite{Goerbig08}) which is taken as origin for all axes.}
    \label{fig:structure}
\end{figure}

At temperatures below the metal-insulator transition, transport and optical investigations
\cite{Dressel94,Clauss10,Ivek11} reveal the development of an energy gap
that is also found in the band structure calculated by density functional theory,\cite{Kino06,Alemany12,Peterseim16a} as illustrated in Fig.~\ref{fig:Bandstructure} of the Appendix~\ref{sec:Bandstructure}.
It is not clear, however, what really happens when uniaxial or hydrostatic pressure is applied.
From magnetotransport measurements\cite{Tajima06,Tajima12c} it was concluded that at high pressure \aeti\ is best characterize as a semiconductor with an extremely narrow energy gap of less than 1~meV. Band structure calculations indicate that the bands actually touch each other at the Fermi energy,\cite{Kino06,Alemany12} supporting previous suggestions of Suzumura and collaborator.\cite{Kondo05,Katayama06,Kobayashi07}
They predicted a zero-gap state under high pressure, where --~in contrast to graphene~--
the Dirac  points do not occur at high-symmetry points and can be tuned by pressure.
%but at $(0.602\pi,-0.353\pi)$ and a second point inversion-sym\-met\-ric to the first; from there it shifts as the crystal is compressed further.
Although the Dirac cone is  anisotropic and tilted,\cite{Tajima12c,Kajita14}
as illustrated in Fig.~\ref{fig:structure}(c), the band dispersion might allow the observation of massless fermions.
This puts \aeti\ in a series of several other crystalline bulk
materials with similar electronic properties
often called three-dimensional Dirac semimetals that have attracted enormous attention in recent years.\cite{Wehling14}

\section{Characterization}

The structure of \aeti\ shown in Fig.~\ref{fig:structure} consists
of planes of BEDT-TTF molecules alternating with layers of I$_{3}^-$
ions. Within the plane the BEDT-TTF molecules are arranged in two chains
with a heringbone pattern. Chain 1 consists of molecules A and
A$^{\prime}$ which are identical at high temperatures but loose
their inversion symmetry at low
temperatures.\cite{Kakiuchi07,Alemany12} The other two molecules in
the unit cell, molecules B and C, constitute chain~2. It is remarkable
that even at high $T$, the charges on the four entities are not
identical, but highest for molecule B and lowest charge density on molecule C.\cite{Kakiuchi07,Ivek11}
Alemany {\it et al.} realized that the origin of this high-temperature
charge disproportionation lies in the arrangement of the tri-iodine
molecules relative to the BEDT-TTF molecules.\cite{Alemany12}

\begin{figure}[b]
    \centering
        \includegraphics[width=0.8\columnwidth]{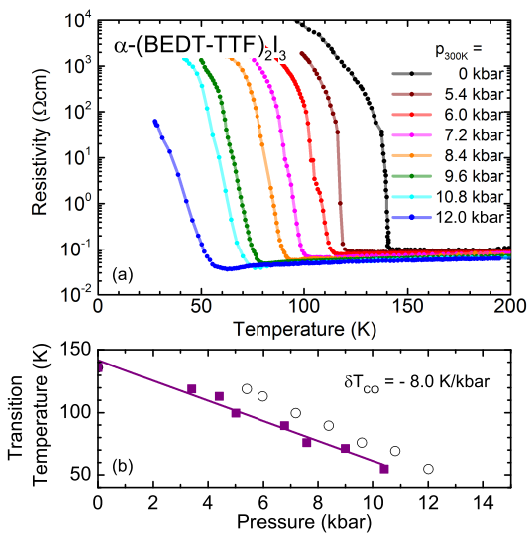}
        \caption{(Color online) (a)~Temperature dependence of the dc resistivity measured within the highly-conducting $ab$-plane. With increasing hydrostatic pressure the transition shifts toward lower temperatures and becomes less abrupt. Panel (b) illustrates the pressure dependence of the metal-insulator transition $T_{\rm CO}$. The open circles refer to the pressure values measured at room temperatures while the solid dots correspond to the corrected pressure values at the actual transition.}
    \label{fig:dc}
\end{figure}
Under ambient conditions \aeti\ can be considered as an electronically quasi-two dimensional system
with an in-plane conductivity ratio of 2 and an out-of-plane
anisotropy of more than 1000. At 135~K it undergoes a metal to
insulator transition,\cite{Bender84b,Dressel94} as seen in Fig.~\ref{fig:dc}(a).
The appearance of a
second Pake doublet in $^{13}$C-NMR\cite{Takano01}, the splitting of
charge sensitive modes in Raman-\cite{Wojciechowski03,Yakushi12} and
IR-spectroscopy,\cite{Moldenhauer93,Dressel04,Ivek11} and careful
x-ray diffraction measurements\cite{Kakiuchi07} reveal a
significant enhancement in the charge disproportionation.
Strong optical nonlinearity and second harmonic generation evidence
that the inversion symmetry between the molecules A and A$^{\prime}$
is broken;\cite{Yamamoto08} the symmetry class of the
crystal changes from P$\overline{1}$ to P1 at low temperatures.
Now, all four molecules are
crystallographically independent and carry a different amount of charge.
Horizontal stripes are formed
with molecule B still being charge-rich and molecule C
charge-poor, but either molecule A or A$^{\prime}$ being charge
rich/poor. This ambiguity in the charge order leads to
a twinned state and the formation of
domains.\cite{Kakiuchi07,Yamamoto10,Ivek10,Tomic15}

Applying pressure on $\alpha$-(BEDT-TTF)$_{2}$I$_{3}$ gradually
suppresses the metal-insulator
transition;\cite{Kartsovnik85,Schwenk85,Tajima00}
a magnetic field, however, recovers an insulating state.\cite{Kajita92,Ojiro93}
In Fig.~\ref{fig:dc}(a) the temperature dependent dc resistivity is plotted for different
values of applied pressure.\cite{remark3}
Above the transition, the conductivity exhibits a very weak metallic temperature dependence.
Extrapolating our data with $\delta T_{\rm CO}=-8.0$~K/kbar [Fig.~\ref{fig:dc}(b)]
yields the complete absence of gap at $p\approx 16.8$~kbar:\cite{remark1}
the system becomes a zero-gap semiconductor.
This Dirac-like semimetal has been suggested theoretically,\cite{Katayama06}
and the linear dispersion explains the very high mobility
and low charge-carrier density measured at high
pressures.\cite{Tajima07}
Experimentally the existence of massless Dirac fermions
has been concluded from various magneto-transport
measurements,\cite{Kajita14} from the NMR relaxation
rate,\cite{Takano10,Hirata11} specific heat data\cite{Konoike12} and
the temperature dependence of the Nernst effect under
pressure.\cite{Konoike13} Recently evidence was presented that the
high-mobility massless Dirac carriers coexist with low-mobility
massive holes.\cite{Monteverde13}

%Optical spectroscopy has proven to be a powerful method for the investigation of Dirac electrons in graphene\cite{Horng11,Mak12,Basov14}
%and three-dimensional Dirac systems.\cite{Chen15}
Although the infrared optical properties of
\aeti\ have been measured
repeatedly,\cite{Koch85,Sugano85,Meneghetti86,Yakushi87,
Zelezny90,Dressel94,Clauss10,Ivek10,Ivek11} to our knowledge no
pressure and temperature-dependent optical studies have been
reported by now that allow to follow the suppression of charge order
and explore the properties of the occurring zero-gap state. Using
our recently developed piston pressure cell for infrared
measurements \cite{Beyer15} we thus have performed comprehensive
optical experiments on $\alpha$-(BEDT-TTF)$_{2}$I$_{3}$ that reveal
the influence of pressure on the charge disproportionation and the
suppression of the optical gap towards the zero-gap state.

\section{Vibrational Spectroscopy}
\label{sec:VibrationalSpectroscopy}
\subsection{Ambient Pressure Results}
\label{sec:VibrationalSpectroscopy_ambient}
The BEDT-TTF molecule has three charge sensitive normal modes known
as $\nu_2({\rm a}_g)$, $\nu_3({\rm a}_g)$ and $\nu_{27}({\rm
b}_{1u})$ in the commonly used D$_{2h}$ symmetry (corresponding to
$\nu_3({\rm a})$, $\nu_4({\rm b}_1)$ and $\nu_{22}({\rm b}_{1})$ in
the correct D$_{2}$ symmetry of the ions in the
solid).\cite{Girlando11} For our infrared experiments we focus on
the antisymmetric stretching vibration $\nu_{27}({\rm b}_{1u})$ of
the outer C=C double bonds; its resonance frequency scales with the
charge $\rho$ according to\cite{Yamamoto05,Girlando11}
\begin{equation}
\nu_{27}(\rho)=(1538-140\cdot\rho){\rm cm}^{-1}  \quad .
\label{eq:nu27}
\end{equation}
In order to observe this vibration, the electric field has to be
polarized parallel to the BEDT-TTF molecule, i.e.\ in case of \aeti\
perpendicular to the $ab$-conducting planes. With typical crystals
as thin as 60~$\mu$m, this requires either the use of a microscope
or measurements on powdered samples.

\begin{figure}
    \centering
        \includegraphics[width=0.9\columnwidth]{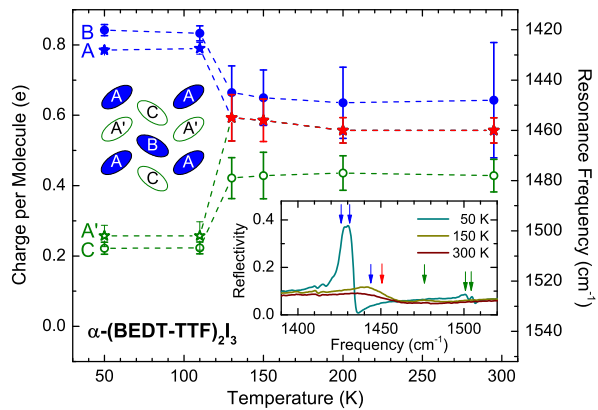}
        \caption{(Color online) Temperature dependence of the charge
distribution of $\alpha$-(BEDT-TTF)$_{2}$I$_{3}$ determined by
infrared reflection measurements with the light polarized
perpendicular to the $ab$-plane.
Following the charge sensitive $\nu_{27}$(b$_{1u}$) mode (right axis),
we find that below the transition temperature of
$T_{\rm CO}=135$~K, the charge per molecule changes drastically for all four sites.
According to Eq.~(\ref{eq:nu27}) we can quantitatively estimate the charge imbalance
(left axis)
and suggest an arrangement as sketched in the left inset
according to the annotation of Kakiuchi \textit{et al.}\cite{Kakiuchi07}
The right inset shows the raw data of reflectivity in the frequency range
of the $\nu_{27}$-vibrations.
The positions of the vibrational modes are illustrated by the
colored arrows, corresponding to the dots in the main frame.
At ambient conditions, blue color indicates the low-frequency vibration of the B molecule,
green the C molecule at high frequencies, and the red arrows the molecules of type A located
in between. The CO transition is clearly visible
by the splitting of the peak: now two features are present each one with a double-peak structure.}
    \label{fig:ambient}
\end{figure}
We have performed ambient-pressure measurements on a single crystal
of \aeti\ utilizing a Bruker Vertex 80v Fourier-transform spectrometer
in combinatation with a Hyperion microscope. Low temperatures were
achieved via a helium-flow cryostat that fits the Cassegrain
objective. The measured reflectivity in the frequency range of the
$\nu_{27}$-vibration is shown in the inset of
Fig.~\ref{fig:ambient}. At elevated temperatures, it consists of two
broad bands.
The very weak band at higher frequencies is due to the
response of the charge poor molecule C, while the information about
the charge on molecules A(A$^{\prime}$) and B is contained in the
stronger band just below 1450~\cm. Sensitive fits with
phenomenological Lorentz and Fano models\cite{Ivek11,Sedlmeier12}
barely allows for a separation of the two contributions, leaving a
significant uncertainty as indicated by the error bars in Fig.~\ref{fig:ambient}.
Below the transition temperature $T_{\rm
CO}$ one very strong band around 1425~\cm\ and a weaker one
just above 1500~\cm\ appear;
as demonstrated in the inset of Fig.~\ref{fig:ambient}, each band contains two contributions.
Based on Eq.~(\ref{eq:nu27}) the lower-frequency modes correspond to $+0.79e$ and $+0.84e$ charge on the BEDT-TTF molecule, and the upper-frequency ones to $+0.25e$ and $+0.22e$.
The measured and fitted resonance frequencies coincide
well with optical results previously
reported,\cite{Yue10,Ivek11,Yakushi12} but the estimates based on
x-ray diffraction\cite{Kakiuchi07} consistently differ by
approximately 10\%.

\subsection{High-Pressure Results}
\label{sec:VibrationalSpectroscopy_highpressure}
For experiments under hydrostatic pressure we first studied a
powdered sample as described in Ref.~\onlinecite{Beyer15} in full
detail. The pressed pellet is put right behind the wedged diamond
window of a copper-beryllium cell filled with Daphne 7373 silicone
oil as pressure-transmitting medium. The cell is attached to a
cold-finger helium cryostat that enables us to reach temperatures as
low as 6~K. In order to adjust for the pressure loss on cooling, we
have performed {\it in-situ} calibration measurements by ruby
flourescence and Manganin wire. Except otherwise stated, only the
actual pressure at any given temperatures is noted throughout the
paper. The infrared reflectivity was measured by employing a Bruker
IFS 66v/S Fourier transform spectrometer.

\begin{figure}
    \centering
        \includegraphics[width=1.0\columnwidth]{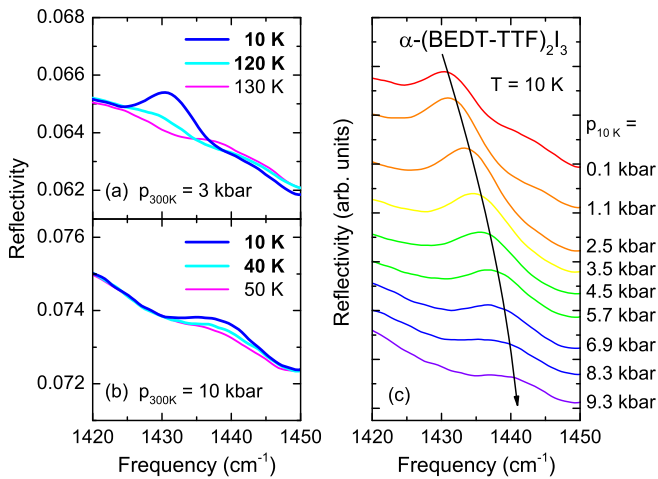}
        \caption{(Color online) Reflection spectra of
$\alpha$-(BEDT-TTF)$_{2}$I$_{3}$ powder measured under pressure,
shown in the frequency range of the $\nu_{27}({\rm b}_{1u})$
vibration. (a)~An applied pressure of 3~kbar reduces the transition
temperature by approximately 15~K. For $T < T_{\rm CO}$ the
reflectivity band clearly raises in intensity. (b)~At higher
pressures, the band correspondingly appears at even lower
temperatures and higher frequency. Because the charge per molecule
is smaller, the feature is also less pronounced. (c)~The pressure
dependence of $\nu_{27}$ vibration at $T=10$~K. With increasing
pressure the resonance frequency shifts up while the strength of
the mode is reduced. The curves are shifted for clarity reasons.}
    \label{fig:v27powerderspectra}
\end{figure}
In Fig.~\ref{fig:v27powerderspectra}(a) we plot data taken at $p_{\rm
300K}=3$~kbar in the frequency range around the strong
$\nu_{27}$-band. Due to the applied pressure, the transition
temperature lies between 130 K and 120 K.
%, cf.\ Fig.~\ref{fig:dc}.
The band around 1427~\cm\
can be observed only in the charge-ordered state. Right below
$T_{\rm CO}$ it is still weak but grows stronger with lower $T$.
While the strength of the band changes with $T$, the
resonance frequency in the insulating state is basically temperature
independent.
At the highest pressure of $p_{\rm 300K}=10$~kbar,
$T_{\rm CO}$ is suppressed to $40-50$~K, and the
resonance frequency of the $\nu_{27}({\rm b}_{1u})$ vibration is
shifted to higher frequencies; as shown in Fig.~\ref{fig:v27powerderspectra}(b)
the reflectivity band becomes much weaker in intensity.
To follow the pressure dependence more systematically, Fig.
\ref{fig:v27powerderspectra}(c) displays the reflectivity spectra of
$\alpha$-(BEDT-TTF)$_{2}$I$_{3}$ for various pressure values
recorded at $T=10$~K. The resonance frequency shifts continuously to
higher frequencies with increasing pressure, and it becomes
significantly weaker. The broadening might to some part be due to
inhomogeneities in the pressure distribution that is not uncommon in
high-pressure powder experiments.

For a quantitative analysis we have to take into account that the
recorded signal $R_{\rm sd}$ is reflected off the diamond/sample-interface
and that in a pressed pellet the in-plane conductivity
contributes considerably.\cite{Beyer15}
Without using a Kramers-Kronig analysis, we directly fit the
observed reflectivity by Fresnel's equations and
modeling the vibrational bands with Fano resonances.\cite{DresselGruner02,Sedlmeier12}
Since the system is over-parametrized, we always
made sure that the parameter variation is kept to a minimum
when gradually changing $p$ and $T$.
The findings are complemented by the ambient-pres\-sure $c$-axis reflectivity
measurement analyzed above in Sec.~\ref{sec:VibrationalSpectroscopy_ambient}, where  the results are displayed in
Fig.~\ref{fig:ambient}. In the metallic state the charge
disproportionation is small: the difference between the
charge-rich molecule B and the charge-poor molecule C only amounts
to roughly $0.2e$ to $0.3e$  This rises abruptly below $T_{\rm CO}$: in
the CO state the charge-rich molecules B and A
differ by approximately $0.6e$ compared to the charge-poor molecules A$^{\prime}$ and C.

\begin{figure}
    \centering
        \includegraphics[width=0.8\columnwidth]{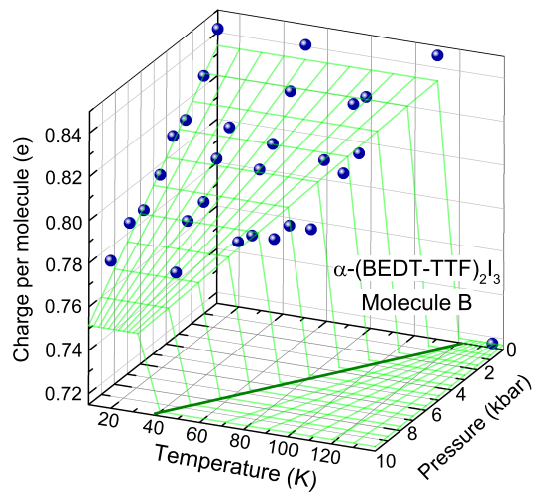}
        \caption{(Color online) The temperature and pressure dependence
        of the charge located on charge-rich molecule B  as determined from the position of the $\nu_{27}({\rm b}_{1u})$ molecular vibrations of \aeti. The thick solid green line indicates the phase boundary at which charge disproportionation occurs.}
    \label{fig:rho_T_p}
\end{figure}
In our pressure-dependent measurements we can only observe the
response of the two charge-rich molecules A and B. The
higher-frequency analogues, i.e.\ the charge-poor sites are much
weaker (cf.\ inset of Fig.~\ref{fig:ambient}) and hard to identify
reliably in the pressurized powder sample.
Fig.~\ref{fig:rho_T_p} summarizes the $p$ and $T$
dependence of the charge on the molecules B, which contains the maximum amount of charge
below $T_{\rm CO}$. Molecule A exhibits a similar behavior. While there is a strong decrease
with $p$, we find that the charge disproportionation does not
vary appreciably with temperature as $T<T_{\rm CO}$.

For a phase transition solely driven by electronic intersite correlations, the charge disproportionation $\Delta\rho$ is a measure of the order parameter and one would expect a gradual increase as $T<T_{\rm CO}$. In our experiments we find a rather abrupt change in charge imbalance at the charge-order phase transition at any pressure: more reminiscent of a first order transition than the second-order behavior observed in the quasi-one-dimensional Fabre salts (TMTTF)$_2X$.\cite{Kohler11,Dressel12} It is tempting to relate this behavior to the structural modifications associated with the coupling to the anions. This was also observed at the charge-order transition in TMTTF salts,\cite{Foury10,Yasin12,Rose13} but probably less dominant and weaker.

\begin{figure}
    \centering
        \includegraphics[width=0.90\columnwidth]{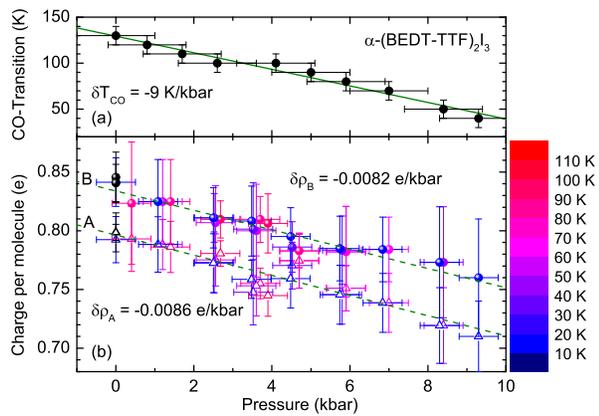}
    \caption{(Color online) Pressure dependence of the transition temperature $T_{\rm CO}$ and the charge disproportionation of \aeti. (a)~The pressure-dependent transition temperature is determined from the optical measurements under pressure. The phase transition to the charge-ordered state $T_{\rm CO}$ shifts linearly to lower temperatures as pressure increases.
(b) The charge per molecule on the charge-rich molecules B (closed circles) and A (open triangle) determined below the transition temperature $T_{\rm CO}$ as a function of pressure. Since in the charge-ordered state the charge is no longer temperature dependent, the actual temperature is coded in the color of the data points. The ambient pressure results taken from Fig.~\ref{fig:ambient} are shown by black symbols. The lines correspond to fits by a linear pressure dependence.}
    \label{fig:pressure}
\end{figure}

Our comprehensive experiments allow us to determine at each individual pressure
the transition temperature $T_{\rm CO}$ at which
charge order starts to become pronounced.
As shown in Fig.~\ref{fig:pressure}(a), the phase boundary decreases
with approximately -9~K/kbar; hence we extra\-polate that at 15~kbar
the charge-order transition should vanish completely. These findings
are in good agreement with our dc measurements plotted in Fig.~\ref{fig:dc}
and previous results by Tajima {\it et al.};\cite{Tajima00,Tajima07}
albeit it should be noted that the later experiments monitor the
metal-insulator transition, while here we actually probe the charge
disproportionation. We conclude a strict coincidence of metal-insulator transition and increase of charge order.

In a next step we can analyze the amount of charge disproportionation reached at low $T$
for different pressure applied.
Fig.~\ref{fig:pressure}(b)
displays the variation of charge on the two molecules A and B with
increasing pressure.\cite{remark2}
The results of the ambient-pressure
measurements are included by the black symbols and coincide well
with the results of the pressure measurement at low $p$ and
low $T$.
For both molecules the charge decreases linearly
by a rate of $\delta\rho_{A} = -0.0086~e$/kbar
and $\delta\rho_{B} = -0.0082~e$/kbar.
Comparing the results presented in both panels,
we conclude that there is a linear relation between
charge disproportionation and transition temperature $T_{\rm CO}$.
Above $14 - 17$~kbar the molecules will carry approximately the same amount of charge at
low temperatures as they do for ambient conditions  in the metallic state
above $T_{\rm CO}$.

The question remains, why the charge disproportionation is leveled off as hydrostatic pressure is applied. In a purely electronic picture\cite{Kino95,Kino96,Seo04} intersite Coulomb repulsion $V$ drives the transition. With pressure the intermolecular distances
are reduced leading to a slight changes of $V$, but most important to
a pronounced enhancement of the orbital overlap,\cite{Kondo05,Kino06}
i.e.\ of the bandwidth $W\propto t$ as discussed in Sec.~\ref{sec:metallicregime}.
In Fig.~\ref{fig:Charge_Vt} of the Appendix~\ref{sec:Tunig_Charg-Order} we plot the amount of charge as a function of effective Coulomb repulsion $V/t$ and find that $\rho_A$ and $\rho_B$  seem to saturate for large Coulomb interaction $V/t$. On the other hand, the interaction of the BEDT-TTF molecules via the ethylene endgroups is supposed to increase with $p$. Alemany {\it et al.} suggested, that
the coupling to the I$_3^-$ anions is crucial for the charge disproportionation.\cite{Alemany12}
At this point it is not clear how the pressure-dependent charge redistribution can be reconciled with this idea. Unfortunately, our temperature and pressure-dependent studies of the ethylene vibrations do not offer any hint in this regard.\cite{remark4}

While we did confine ourselves to the  $\nu_{27}({\rm b}_{1u})$
mode  here, in principle a similar analysis could be performed with other
charge-sensitive molecular vibrations. However, we cannot reach a
comparable sensitivity and accuracy as discussed in Appendix~\ref{sec:EMV-Coupled Vibrations} in more detail.
Instead we compare our findings to pressure and temperature dependent
Raman measurements of the $\nu_{2}$(a$_g$) and $\nu_{3}$(a$_g$) modes.
Due to a lower resolution,  Wojciechowski
\textit{et al.} could not estimate the
charge on each molecule separately but only detect the lowest
and the highest charge and thus estimated the difference $2\delta$
between charge rich and charge poor sites:\cite{Wojciechowski03}  it decreases from
$\delta=0.3e$ at ambient pressure to $0.2e$ at $p=12$~kbar. This
corresponds to a rate of approximately $-0.0083~e$/kbar, in
excellent agreement with our measurements.

\section{Electronic Properties}
\begin{figure}[b]
    \centering
        \includegraphics[width=0.8\columnwidth]{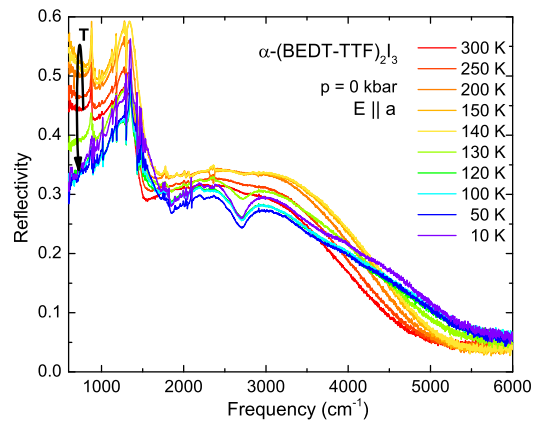}
        \caption{(Color online) Temperature dependence of the optical
reflectivty of \aeti\ measured at ambient pressure with light
polarized along the $a$-axis off the crystal surface outside the
pressure cell. At room temperature a rather well-developed plasma
edge underlines the metallic behavior. As indicated by the black
arrow at the left: when the temperature decreases the low-frequency
reflectivity first rises in the metallic state until it takes a
drastic drop at the charge order transition $T_{\rm CO}=135$~K.}
    \label{fig:refa00}
\end{figure}
\begin{figure}
    \centering
        \includegraphics[width=1.0\columnwidth]{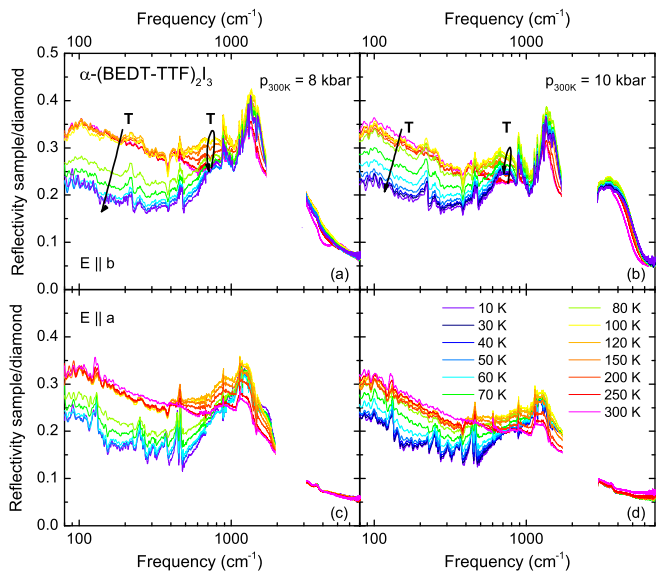}
        \caption{(Color online) The optical reflectivity $R_{\rm sd}$ of \aeti\ versus
frequency measured at different temperatures and pressure values as
indicated. The upper panels display data taken for the highest reflecting
polarization $E\parallel b$, while the lower panels correspond to the
direction $E\parallel a$. The pressure $p_{\rm 300K}=8$ and 10~kbar
refer to the values set at room temperature. The blocked
areas correspond to the spectral range of high diamond absorption.
The black arrows illustrate the temperature behavior, which is
monotonous in the far-infrared but in the mid-infrared spectral
range it exhibits a maximum right above the metal-insulator
transition.}
    \label{fig:reflectivity}
\end{figure}

In order to explore the electrondynamic properties of \aeti\ when
developing charge-order and approaching the zero-gap, the optical
reflectivity off the highly conducting $ab$-plane of a single
crystal has been measured from room temperature down to $T=10$~K
applying pressure up to $p_{\rm 300K}=11$~kbar. Due to the rather
large opening of the diamond window of 2~mm diameter, we can record
data well below 100~\cm.\cite{Beyer15} Examples of the optical reflectance
for different polarizations, pressure values and temperatures
are displayed in Fig.~\ref{fig:reflectivity}. The strong multiphonon absorptions
in diamond between 1700 and 2700~\cm\ prevents any
reasonable signal to be recorded in this range of frequency. Since the
crystal is in optical contact with the window, the reflectivity $R_{\rm sd}$ is
probed at the diamond-sample interface; using the known properties
of diamond, the optical reflectivity is calculated following the
procedure described by Pashkin {\it et al.} \cite{Pashkin06,remark5}

At ambient pressure the low-frequency reflectivity continuously
rises with decreasing $T$ until the metal-insulator
transition is reached at $T_{\rm CO}=135$~K; the metallic frequency
dependence quickly transforms into an insulating behavior that
remains unchanged below approximately 120~K. As demonstrated in Fig.~\ref{fig:refa00}
the well-pronounced plasma edge then transforms into a gradual
decrease of reflectivity with frequency. Detailed discussions of the
ambient-pressure optical properties have been reported previously by several
groups.\cite{Koch85,Sugano85,Meneghetti86,Yakushi87,Zelezny90,Dressel94,Clauss10,Ivek10,Ivek11}

In Fig.~\ref{fig:reflectivity} the optical reflectivity of \aeti\ is
displayed in the far and mid infrared (80 to 8000~\cm) for both
polarizations parallel and perpendicular to the $a$-axis; the plots
correspond to the raw data taken off the diamond-to-sample interface
at different temperatures as indicated.
When a pressure of 8~kbar is applied,
in the spectral range above 500~\cm\ the $T$
dependence is similar to the ambient pressure response,
while the far-infrared reflectivity stays independent
down to the metal-insulator transition.
When charge order sets in at $T=80$~K
the reflectivity decreases abruptly and significantly before it
approaches a constant value below 60~K, as seen from
Fig.~\ref{fig:reflectivity}(a) and (c).

As the applied pressure increases to 10~kbar and more,
the changes with $T$ become more gradual and
less extensive; the transition shifts to around 50~K
and becomes broad.
For the polarization $E\parallel a$
the overall $R_{\rm sd}(\omega)$  is slightly lower, but the temperature
behavior is similar.
In Fig.~\ref{fig:reflectivity}(b) we can see that
there is still a crossing point around 600~\cm\ indicating that
spectral weight shifts to the mid-infrared as the temperature is
reduced.
For $E\parallel a$ the mid-infrared band around 1500~\cm\
is almost absent at elevated temperatures; it grows only when $T$
drops below 200~K.
Even more interesting is the reduction in the
far-infrared reflectivity for $T<80$~K observed in both
polarizations.
This implies a drastic depletion  of charge carriers
in the Drude term.
%In contrast to the ambient and low-pressure behavior, however, the material still remains metallic.

When we compare the low-$T$ reflectance at different
pressure values, we notice an increase of the far-infrared
reflectivity as pressure rises to 10~kbar, but then it comes to a halt.
This indicates a growth in spectral
weight with pressure and a rise of the low-frequency conductivity as
will be discussed in more detail below.

\subsection{Metallic Regime}
\label{sec:metallicregime}
\begin{figure}[b]
    \centering
        \includegraphics[width=0.8\columnwidth]{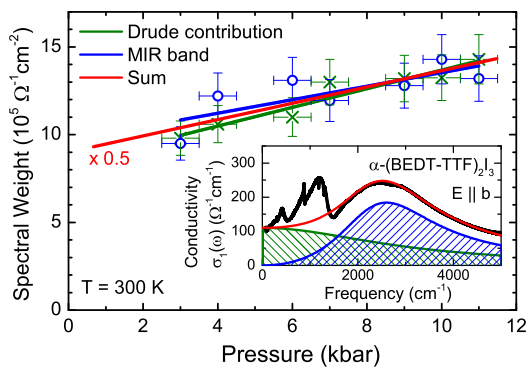}
        \caption{(Color online) Pressure dependence of the spectral weight
of \aeti\ obtained by Drude-Lorentz fits of the room-temperature
optical data (black line) as sketched in the inset: the green line and crosses correspond to the Drude component, while blue line and open circle refers to the
mid-infrared band; the red line is the sum of both.
With increasing pressure both contributions become significantly larger.
Here we present the results for the polarization
$E\parallel b$; for the $a$-direction the behavior is similar, but the absolute values
smaller by a factor of 2.}
    \label{fig:spectralweight}
\end{figure}
In order to study the influence of pressure on the bandwidth and
carrier density in more detail, we have fitted the
overall optical conductivity of \aeti\ by a sum of Drude and Lorentz
terms, disregarding all vibrational features and emv-coupled modes.
Previous bandstructure calculations\cite{Emge86,Ishibashi06,Kino06,Kondo09,Mori10,Alemany12} yield a semimetallic behavior at ambient conditions: the Fermi surface intersects the valence band and conduction band leading to electron and small hole pockets.
Our DFT calculations\cite{Peterseim16a} sketched in the Appendix~\ref{sec:Bandstructure}, however, evidence only one kind of carriers at $T=300$~K (Fig.~\ref{fig:Bandstructure}).
Since also the experimental results do not provide any information that allows us to
discriminate two Drude contributions according to two separate bands,
we restrained ourselves to the simples possible model of one Drude and one Lorentz term.
Although this might be a crude and simplified approach, it
allows us to disentangle the free charge-carrier contribution
from localized electrons, interband transition, etc.\ centered in the
mid-infrared spectral range.\cite{DresselGruner02,Dressel09}
Both contributions carry about the same spectral weight $\int\sigma(\omega){\rm d}\omega = \omega_p^2/8$, with $\omega_p$ the plasma frequency.
The overall properties are similar for both polarizations, however, the gross spectral weight parallel to the chains is only half compared to $E\parallel b$.

When hydrostatic pressure is applied, the spectral weight increases in a linear
fashion as shown in Fig.~\ref{fig:spectralweight}.
The enhancement of the Drude component is more pronounced compared to the
mid-infrared band; in total we find a rise of about 50\%\ at 12~kbar, similar for both polarizations.
In a first approach this behavior is explained by the enlargement of the bandwidth $W$
as the lattice is compressed.
DFT calculations also yield an increase of $W$ of 27\%\
at high pressure.\cite{Kino06,Alemany12}.
The $p$ dependence of the dc resistivity (Fig.~\ref{fig:dc}) confirms this conclusion that \aeti\ becomes more metallic with pressure.
We have to keep in mind, however, that $\omega_p^2$ is proportional to the ratio of
carrier density and mass; a reduction in correlations is commonly ascribed to a decrease in the effective mass.
It is interesting to note, that in the case of the charge-fluctuating
metal $\alpha$-(BEDT-TTF)$_2$KHg(SCN)$_4$ spectral weight moved from
the mid-infrared band to the Drude part when hydrostatic pressure of
up to 10~kbar is applied. This was interpreted as a reduction of the
effective correlations and shift towards metallic
behavior.\cite{Drichko06}

\begin{figure}[b]
    \centering
        \includegraphics[width=0.9\columnwidth]{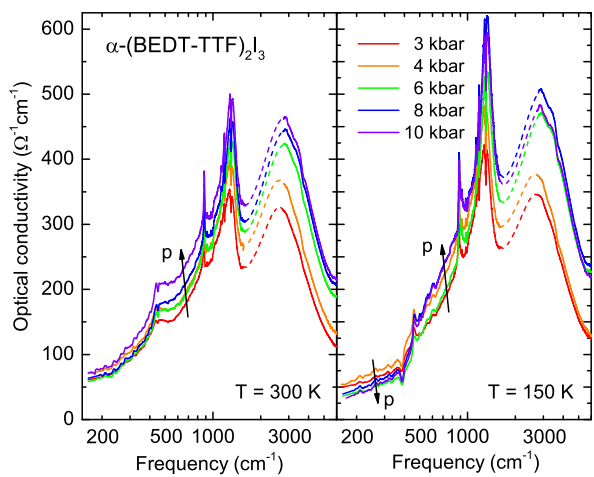}
        \caption{(Color online) The frequency-dependent conductivity of
\aeti\ measured along the $b$ direction at pressure at $T=300$ and
$150$~K. The dashed lines indicate the range of interpolation due to
high diamond absorption.}
    \label{fig:conductivity}
\end{figure}
This pressure behavior also becomes obvious from the conductivity
spectra at $T=300$ plotted in Fig.~\ref{fig:conductivity}. Cooling
down from room temperature just above the metal-insulator transition
($T= 150$~K), the overall conductivity of \aeti\ increases by
approximately 10~\%, in accord with previous ambient-pressure
measurements and dc transport.\cite{Dressel94} While
the effects on pressure is rather similar around the
mid-infrared peak at 2500~\cm,
the pressure dependence becomes more subtle
in the far infrared range where a crossover can be identified. This
behavior is more pronounced when we cool down further into the
zero-gap regime.

\subsection{Narrow-Gap Regime}
The optical properties below the metal-insulator transition of
\aeti\ are governed by two contributions:
first, the free-carrier absorption, which freezes out as
$T$  is reduced, as demonstrated in Fig.~\ref{fig:refa00};
second, by excitations across the gap,
which is supposed to gradually close as pressure is applied.
At ambient pressure the optical gap $\Delta$
is clearly seen in the conductivity spectra around 600~\cm,
similar for both polarizations.\cite{Clauss10,Ivek11}
Our bandstructure calculations plotted in Fig.~\ref{fig:Bandstructure}(b) show
that only 21~meV separate the bands from each other; the direct gap amounts to 55~meV.\cite{Peterseim16a} This is in perfect agreement with the values of
30 and 60~meV, respectively, given by Alemany {\it et al.}\cite{Alemany12}
Despite several high-pressure resistivity or Hall effect investigations
down to low temperature\cite{Tajima00,Tajima06,Tajima12b}
nothing is known about the evolution of the transport gap with increasing $p$.

\begin{figure}[h]
    \centering
        \includegraphics[width=0.8\columnwidth]{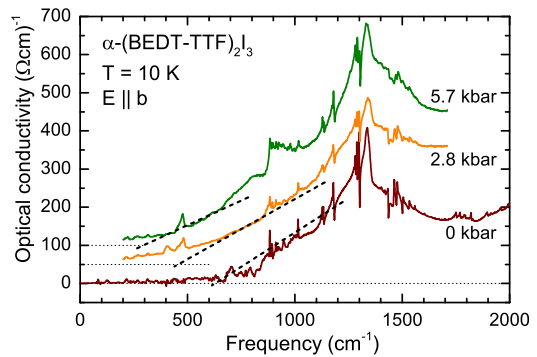}
        \caption{(Color online) Low-temperature conductivity of \aeti\ as a
function of frequency for $E\parallel b$. By extrapolating
the low-frequency behavior to zero (dashed lines), the optical gap
can be extracted, as shown here for $p=0$ and 2.8 and 5.7~kbar; the curves are displaced by $50~(\Omega {\rm cm})^{-1}$. }
    \label{fig:pressuregap}
\end{figure}
In Fig.~\ref{fig:pressuregap} we plot $\sigma(\omega)$ of \aeti\
at $T=10$~K for different amounts of hydrostatic pressure.
Similar to the high-$T$ behavior, the overall conductivity increases;
but now it extends down to the far-infrared spectral range.
Since we can unambiguously determine the optical gap for ambient pressure
by linear extrapolation of $\sigma(\omega)$ to low frequencies,
a similar procedure is applied to extract the optical gap $\Delta$ for different pressure values, as illustrated by the dashed lines in Fig.~\ref{fig:pressuregap}.
A rather similar behavior is observed for the polarization $E\parallel a$.
We want to mention that the determination is tainted with a
considerable uncertainty due to the enormous challenge of the low-frequency
and high pressure measurements, due to the ambiguity of the data
analysis via Drude-Lorentz-fits and Kramers-Kronig analysis, due to
the uncertainty of extrapolation etc. Also note that the optical conductivity
does not actually drop to zero within the accessible frequency range,
and this background seems to become more pronounced as pressure increases;
it will be discussed in the preceding Sec.~\ref{sec:DiracCone}.

\begin{figure}
    \centering
        \includegraphics[width=0.84\columnwidth]{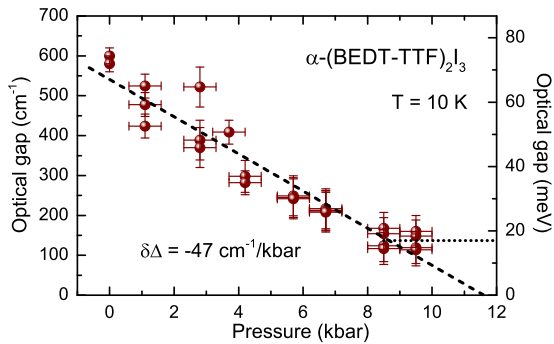}
        \caption{(Color online) Pressure dependence of the energy gap
$\Delta(p)$ of \aeti\ extracted from low-temperature optical
reflectivity measurements. The multiple data points at one particular
pressure value correspond to the analysis for different polarization
directions and repeated measurements. With increasing pressure the
extracted gap value drops linearly to zero at approximately 11-12~kbar
(dashed line); however, the data indicate that the decrease gets to
a halt with a
small but constant optical gap of approximately 16~meV, indicated by
the dotted line.}
    \label{fig:opticalgap}
\end{figure}
Nevertheless, Fig.~\ref{fig:opticalgap} summarizes the pressure dependence of
the energy gap extracted from our optical measurements on \aeti\
down to low temperatures and frequencies.
The data evidence a linear decrease of the optical gap $\Delta(p)$
with increasing pressure of approximately -47~\cm/kbar, i.e. $\delta \Delta = -6$~meV/kbar.
Despite the error bars, for the highest accessible pressure values
we cannot see any further change above the actual low-temperature pressure of 9~kbar.
This might taken as an indication that the closing of the gap actually come to a
halt with $\Delta\approx 16$~meV. At a first glance this observation seems in
contradiction to the metallic behavior reported from dc resistivity
measurements;\cite{Tajima00,Tajima12c} however, we have to keep in mind that
optical experiments only probe direct transitions ($\Delta
q=0$),\cite{DresselGruner02} while temperature-dependent transport
measures the smallest distance between two bands, i.e. indirect
gaps. From the optical point of view \aeti\ actually becomes a
semimetal and not a real zero-gap state. We certainly have to
extend our optical experiments to higher pressure and lower
frequencies in order to give a final answer to this problem, because
calculations of the electronic structure based on the high-pressure
x-ray scattering results\cite{Kondo05} indicate that the
two bands in of \aeti\ actually touch at the Fermi energy.\cite{Kino06,Alemany12} Here calculations of the optical conductivity for low $T$ and different pressure would be helpful, as done for the ambient condictions.\cite{Peterseim16a}

\subsection{Dirac Cone and Zero-Gap State}
\label{sec:DiracCone}
\begin{figure}[h]
    \centering
        \includegraphics[width=0.8\columnwidth]{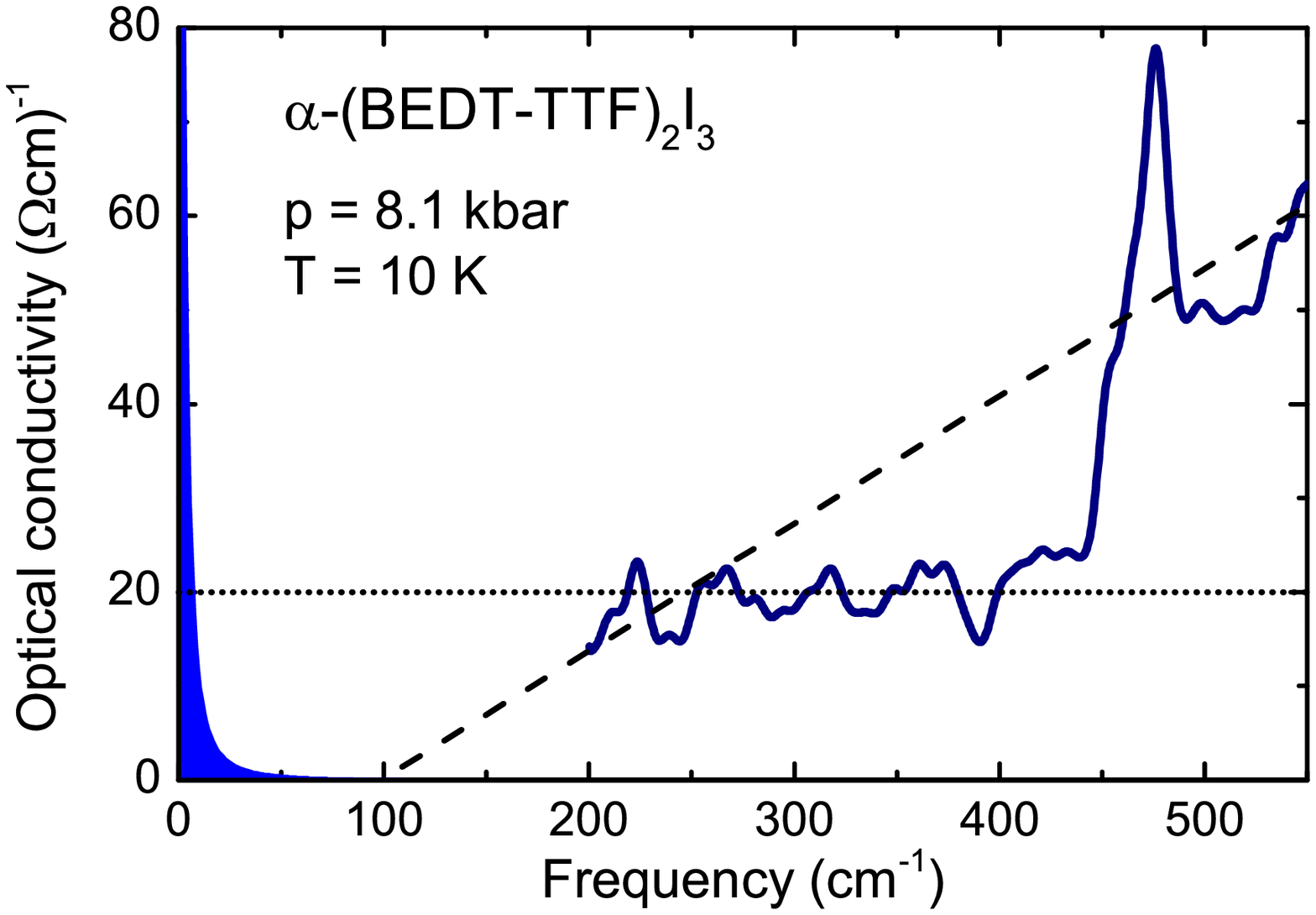}
    \caption{(Color online) Low-frequency conductivity of \aeti\ for $E\parallel b$, $p=8.1$~kbar and $T=10$~K.
The dashed lines illustrates the attempt to estimate the optical gap by a linear extrapolation of the higher-frequency conductivity. Below 450~\cm\ there seems to remain a constant conductivity of $\sigma = 20~(\Omega {\rm cm})^{-1}$ for the highest-pressure measurement (dotted line).
The dc conductivity is indicated by a narrow Drude peak with a roll off at 2~\cm.}
\label{fig:Lowfreqcond}
\end{figure}

Above we saw that the low-frequency optical conductivity of \aeti\
does not completely vanish %at low temperatures
when high pressure is applied.
Even for lowest temperatures there seems to remain a constant background
of approximately $\sigma=20~(\Omega{\rm cm})^{-1}$ as demonstrated in Fig.~\ref{fig:Lowfreqcond}.
Is this an indication that the bands start to overlap
and \aeti\ becomes a semimetal? Or do the bands touch
and we see fingerprints of the Dirac electrons?

In order to clarify  whether this background stems from normal
massive carriers or from massless Dirac electrons, let us compare
the value with transport measurements.
From high-pressure data of magnetoresistance and Hall coefficient
$R_H$, it is known\cite{Tajima00,Tajima06} that at low temperatures
($T=10$~K) \aeti\ contains an equal amount of electrons and holes
with a density of only $3 \times 10^{16}~{\rm cm}^{-3}$ and an
extremely high mobility of $3\times10^4~{\rm cm}^{2}$/Vs. The
conductivity estimated by $\sigma= n e\mu\approx 150~(\Omega{\rm
cm})^{-1}$ is slightly above the value of approximately
$50-100~(\Omega{\rm cm})^{-1}$ obtained from dc
measurements.\cite{Tajima07}
This implies that a very narrow Drude contribution is present.
Using $\mu=e\tau/m$, with $m$ the free
electron mass, the scattering rate can be estimated
to $\tau^{-1}=6\times 10^{10}~{\rm s}^{-1}$, which corresponds to
2~\cm; the respective Drude term is sketched in Fig.~\ref{fig:Lowfreqcond}.
In other words, if the low-$T$ dc conductivity was caused by
massive carriers, the Drude roll-off would fall far below the
frequency range accessible to us in the present experiments.
Previous microwave experiments between 10 and 600 GHz (0.3 - 18~\cm)
did find a rather strong frequency dependence,\cite{Dressel94}
but have been restricted to ambient pressure;
probing the low-temperatures microwave response as a function of frequency at high-pressure
is not possible at present time.

Some electronic-structure calculations indicate semimetallic bands of \aeti\ under ambient con\-ditions;\cite{Kino06,Mori10,Alemany12}
at high pressure this might also extend to low temperatures.
In a first approximation the electronic properties of a semimetal can be treated by two independent conduction channels --~one for electrons and one for holes~-- leading to an optical response that is described by the sum of two Drude terms.\cite{DresselGruner02}  If the corresponding relaxation rates differ strongly, both  contributions can be discriminated;\cite{Wu10,Barisic10,Homes15} in general, however,
comparable scattering processes and interband scattering prevent an unambiguous identification and clear-cut separation. In the present case of \aeti\ we do not observe a Drude-like response at low temperatures and cannot conclude on the iussue of one or two-band electronic transport.

Do we see the optical response of the Dirac cone? Based on their semi-empirical investigations of the electronic structure,
Kata\-yama, Kobayashi, and Suz\-umura\cite{Katayama06,Kobayashi07}
suggested that a two-di\-men\-sional an\-iso\-tropic Dirac cone dispersion occurs
in \aeti\ at high pressure,
in analogy to the linear dispersion of the energy bands of graphene,
although the situation is distinct as the states are not protected by topology, the Dirac cone is tilted and the zero-gap states can be tuned by pressure.\cite{Tajima12c,Kajita14}
%In recent years several crystalline bulk materials have been reported with similar electronic properties, often called three-dimensional Dirac semimetals.
Very recently,
Suzumura {\it et al.} theoretically examined the dynamical conductivity of the massless
electrons in the tilted Dirac cone,\cite{Suzumura14a,Suzumura15}
i.e.\ with different velocities for the first and second band.
They found that the behavior deviates from the case of graphene as
intraband excitations are enhanced by the tilting compared to that
of the interband excitations.

Albeit \aeti\ was the first bulk material were Dirac electrons have been suggested, electronically it can be considered strictly two-dimensional and thus more similar to graphene (cf.\ Appendix~\ref{sec:Dirac_Cone}).
Hence the interband optical conductivity per layer should be an universal constant\cite{Gusynin06}
\begin{equation}
G(\omega) = N\frac{\pi}{8}G_0 \quad , \label{eq:Fermivelocity}
\end{equation}
where
$G_0=2e^2/h=7.748\times 10^{-5}~\Omega^{-1}$ is the conductance
quantum, $h=2\pi \hbar = 6.626\times 10^{-34}$~Js Planck's
constant, and  $N$ is the number of non-degenerate cones ($N = 2$ for graphene
while $N = 4$ in the case of \aeti). The skin depth, $\delta = c\left(2\pi
\sigma \omega\right)^{-1/2} $, is above 2~$\mu$m for any $\omega$ in our
sample, hence the specific bulk conductivity due to the Dirac particles
should be below $1~(\Omega{\rm cm})^{-1}$. This is, however, not
the case for any frequency. In other words, the contribution of the
Dirac electrons is masked by other channels of conduction.

Our observation is in full accord with the findings of
Monteverde {\it et al.} who concluded from magnetotransport measurements
under hydrostatic pressure  that in \aeti\
low-mobility massive holes coexist with the highly-mobile massless Dirac carriers.
While transport measurements sum
over all carriers, optical measurements are energy selective and
thus could provide additional important information in this regard.
Unfortunately, the optical conductivity presented in Fig.~\ref{fig:Lowfreqcond}
does not reach high-enough pressure and low-enough frequencies to
make definite statements on the true zero-gap state in \aeti.
Experiments in the THz range and at higher pressure are required
in order to disentangle the various contributions to the optical conductivity
--~massless and massive carriers as well as interband and intraband excitations~--
and determine their dynamical properties.

\section{Conclusion}
We have measured the optical properties of the organic conductor
\aeti\ under hydrostatic pressure down to low temperatures and low
frequencies.  At elevated temperatures the metallic response observed by
transport and optics improves with pressure due to the enlarged bandwidth and
enhanced carrier density.
As pressure rises, the metal-insulator transition is suppressed by -8~K/kbar,
this coincides with the increasing charge disporportionation at
$T_{\rm CO}$ for which the shift was independently estimate to -9~K/kbar by our
infrared measurements. In our vibrational spectra we see how the charge imbalance
$\delta \rho$ decreases linearly with pressure by $\Delta\rho = 0.017~e$/kar;
above approximately 14~kbar the charge
per molecules reaches the values known from above the charge
ordering transition.

Also at low temperatures the metallic properties become enhanced by pressure
and the gap between conduction and valence band is strongly suppressed. The optical gap decreases by -6~meV/kbar up to 9~kbar.
Since we do not observe a Drude-like response for the maximum pressure reached,
we characterize \aeti\ as a semiconductor with an extremely narrow gap;
we cannot see the bands touching.
For the highest pressure we find that a constant low-frequency conductivity of
$20~(\Omega {\rm cm})^{-1}$ remains at small temperatures. Although this is reminiscent of linear dispersion in two-dimensional Dirac systems, the absolute value seems to be
too high. We discuss how massive Drude and massless Dirac charge carriers contribute to the high-pressure optical response.

\acknowledgements We  thank N. Barisi{\'c}, N. Drichko,
E. Rose, D. Wu and S. Zapf for helpful discussions as well
as G. Untereiner for technical support. Funding by the Deutsche
Forschungsgemeinschaft (DFG) and Deutscher Akademischer
Austauschdienst (DAAD) is acknowledged.

\appendix
\section{Electronic Band Structure}
\label{sec:Bandstructure}
The band structure of \aeti\ was calculated by \textit{ab-initio} density functional theory (DFT) as standardly implemented in the software package Quantum Espresso (Version 4.3.2 and 5.1).\cite{Giannozzi09} We employed a norm-conserving PBE general gradient approximation (GGA) functional \cite{Perdew96} for all atom types, up to a certain level taking into account the exchange correlation as well as the spatial variation of the charge density.
The cut-off energy for the plane waves and electronic density was set to 30~Ry and 120~Ry, respectively. The self-consistent energy calculations were performed on a regularly spaced grid of $8\times8\times4$ grid.\cite{Monkhorst76} Since \aeti\ is metallic at room temperature, a smearing factor of 0.05~Ry was selected. The crystal structures determined from x-ray scattering experiments at room temperature and low temperature were taken from Refs.~\onlinecite{Kakiuchi07} and \onlinecite{Emge86}. They are used without any optimization of the unit cell parameters or the atomic positions.
\begin{figure}[h]
    \centering
    \includegraphics[width=0.8\columnwidth]{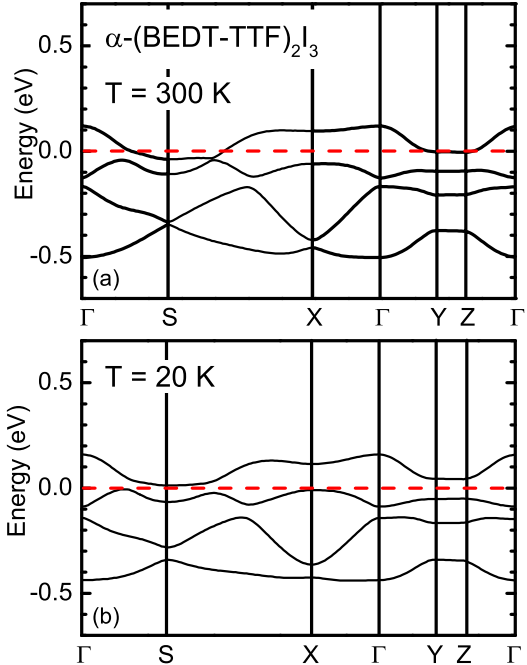}
   \caption{(Color online) Calculated band structure of \aeti\ (a) in the metallic state at  $T=300$~K and (b) in the insulating state at $T=20$~K. The Fermi level $E_F$ is indicated by the red dashed lines. \label{fig:Bandstructure}}
\end{figure}

In Fig.~\ref{fig:Bandstructure} we plot the electronic band structure of \aeti\
along the path $\Gamma (0,0,0)\rightarrow S (-0.5,0.5,0)\rightarrow  X (0.5,0,0) \rightarrow \Gamma (0,0,0) \rightarrow Y (0,0.5,0) \rightarrow \text{Z} (0,0,0.5) \rightarrow \Gamma (0,0,0)$ in units of the triclinic reciprocal lattice vectors.
Since the unit cell contains four molecules, four bands are found at the Fermi energy $E_F$ (depicted by the red dashed line).
The lower two are well separated from the Fermi level whereas the upper band intersects $E_F$. From that, we conclude that \aeti\ is a metal with electrons as major carriers, which agrees with previous calculations\cite{Kino06,Mori10,Alemany12} where also electron pockets were found and additionally small hole pockets. As previously discussed by Alemany {\it et al.}\cite{Alemany12}, the difference can be ascribed to numerical uncertainties in the calculations since the difference are on the meV regime as well as on the (unrelaxed) used crystal structures leading to small deviations.

At $T=20$~K the bands shift and an indirect gap of 21~meV opens, as shown in Fig.~\ref{fig:Bandstructure}(b); optical excitations
see the smallest direct gap of 55~meV. For more details on the calculated optical properties see Ref.~\onlinecite{Peterseim16a}. Very similar results were reported
by Alemany {\it et al.}\cite{Alemany12} We also want to note that the bands exhibit basically no dispersion in the $c$-direction, independent on temperature.  Treating \aeti\ as a two-dimensional metal or narrow-gap semiconductor, respectively, seems to be a rather good choice.

\section{Tuning the Charge-Order }
\label{sec:Tunig_Charg-Order}
In order to illustrate the dependence of the charge order on electronic correlations, we have replotted the molecular charge $\rho_A$ and $\rho_B$ given in Figs.~\ref{fig:rho_T_p} and \ref{fig:pressure} as a function of effective intersite Coulomb repulsion $V/t$ normalized to the ambient pressure value. The conversion from hydrostatic pressure to effective Coulomb interaction $V/t$ was estimated by using the pressure dependent lattice parameters,\cite{Kondo05,Kino06,Kondo09} assuming a slight $r^{-2}$ distance dependence of the electronic interaction and a strong linear increase of the hopping integral $t$ with pressure, known also from comparable organic charge-transfer salts.\cite{Rose13,Jacko13}

\begin{figure}[h]
    \centering
        \includegraphics[width=0.8\columnwidth]{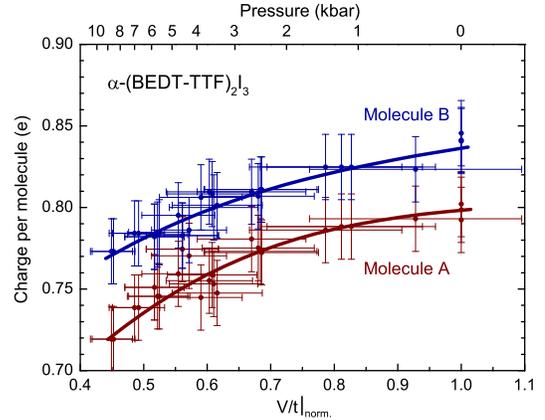}
        \caption{(Color online) Variation of the charge located at molecule A (red dots) and B (blue dots) in the charge-ordered phase $T<T_{\rm CO}$ when the effective intersite Coulomb repulsion $V/t$ increases (lower axis) by applying hydrostatic pressure $p$ (upper axis). The charge per molecule $\rho_A$ and $\rho_B$ is determined from the shift of the $\nu_{27}({\rm b}_{1u})$ vibrational modes. The lines are guides to the eye.}
    \label{fig:Charge_Vt}
\end{figure}
The uncertainty in determining the charge on the molecules A and B from the shift of the $\nu_{27}({\rm b}_{1u})$ molecular vibrations, in measuring the actually applied pressure at low temperatures and in evaluating the effective Coulomb repulsion from the hydrostatic pressure leads to large error bars. Nevertheless, by summarizing the data from various temperatures and pressure runs, we can identify a similar behavior for both molecules.
For $V\rightarrow 0$ it extrapolates to the values of 0.64 and $0.56e$ determined in the metallic state (Fig.~\ref{fig:ambient}).
With increasing $V/t$ the molecular charge density seems to saturate at approximately the maximum value obtained at ambient pressure and low temperatures.

\section{EMV-Coupled Vibrations}
\label{sec:EMV-Coupled Vibrations}
\begin{figure}[b]
    \centering
        \includegraphics[width=0.45\columnwidth]{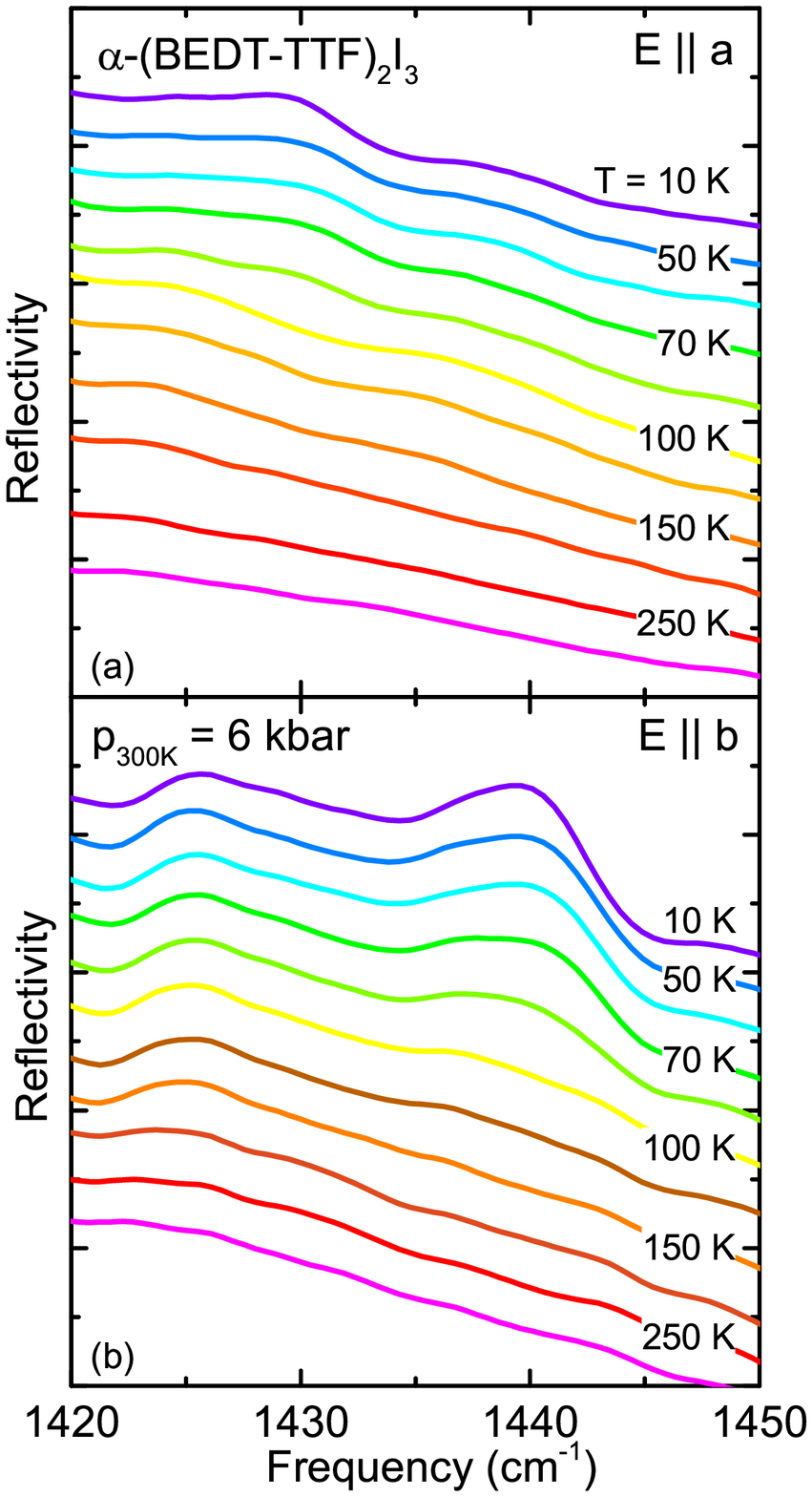}\hspace*{3mm}
                \includegraphics[width=0.45\columnwidth]{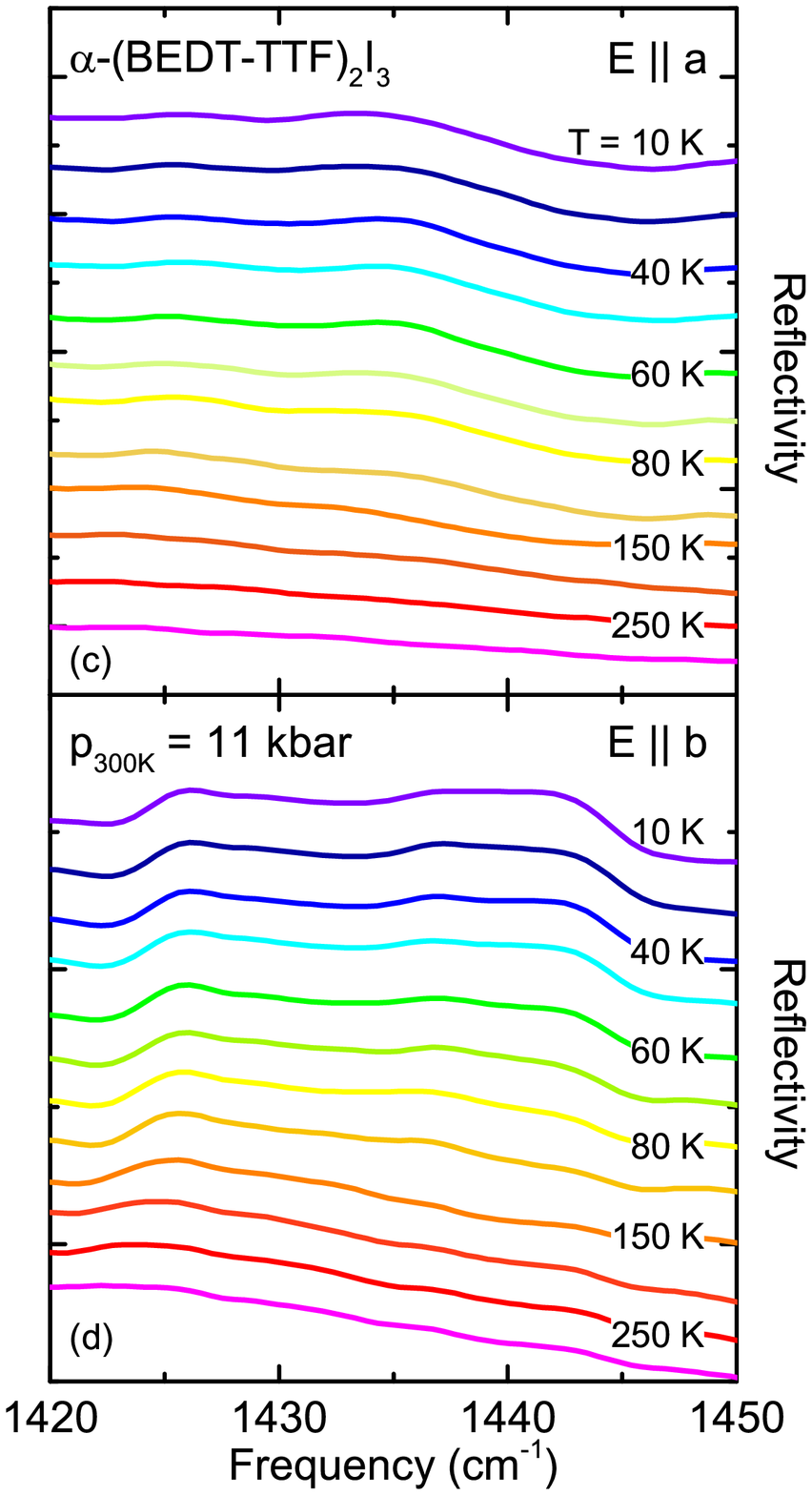}
        \caption{(Color online) Frequency-dependent reflectivity of
$\alpha$-(BEDT-TTF)$_{2}$I$_{3}$ measured for the polarization (a,c) $E \parallel a$ and
(b,c) $E\parallel b$ at different temperatures as indicated. Vibrational features develop at low temperatures. They shift to higher frequencies as the pressure is increased from $p_{\rm 300K}=6$~kbar to 11 kbar.}
    \label{fig:nu3-mode}
\end{figure}
In the Sec.~\ref{sec:VibrationalSpectroscopy} we analyzed only the asymmetric $\nu_{27}$(b$_{1u}$) mode obtained from the spectra taken at pressed pellets.
This can be complemented by reflection experiments off the $ab$-plane (see below)
where fingerprints of the charge-sensitive modes $\nu_{2}$(a$_g$) and $\nu_{3}$(a$_g$)
become visible in the same range of frequency. These are symmetric vibrations of the two C=C bonds that become infrared active via electron-molecular vibrational (emv) coupling.\cite{Dressel04,Girlando11} Their dependence on the charge located on the BEDT-TTF molecule is comparable to the one of the $\nu_{27}$(b$_{1u}$)  mode.
In Fig.~\ref{fig:nu3-mode} we plot the optical reflectivity for different polarizations and temperatures taken at a pressure values of 6 and 11~kbar. The vibrational features become more pronounced at low temperaturs; they shift to higher frequencies as pressure increases indicating the significant reduction of charge. Due to the emv coupling and broad spectral feature, we refrain from a quantitative analysis of the temperature and pressure dependence of the $\nu_{2}$(a$_g$) and $\nu_{3}$(a$_g$) modes and refer to more sensitive Raman experiments.\cite{Wojciechowski03}

\section{Dirac Cone}
\label{sec:Dirac_Cone}
\begin{figure}[h]
	\centering
		\includegraphics[width=0.8\columnwidth]{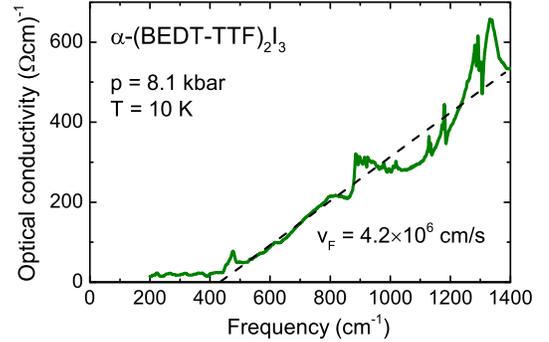}
	\caption{(Color online) Low-frequency conductivity of \aeti\ for $E\parallel b$, $p=8.1$~kbar and $T=10$~K.
The dashed line corresponds to a linear fit of the region between 500 and 1400~\cm.}
	\label{fig:Fermivelocity}
\end{figure}

By looking at the frequency dependent conductivity of \aeti\ at low $T$ and high $p$,
plotted in Fig.~\ref{fig:Fermivelocity}, one is puzzled by the large range with a linear increase: between 500 and 1400~\cm\ the conductivity follows $\sigma(\omega)\propto \omega$.
Recently such a linear frequency dependence in the optical conductivity of  quasicrystals,\cite{Timusk13} ZrTe$_5$,\cite{Chen15} and Cd$_3$As$_3$\cite{Neubauer16}
was considered the hallmark for Dirac physics in three dimensions.
The arguments are based on calculations\cite{Hosur12,Bacsi13} for the interband optical response of $d$-dimensional Dirac systems, where a power-law frequency dependence $\sigma(\omega)\propto\omega^{(d-2)/z}$ was found, with the exponent $z$ defined by the energy dispersion $E(k)\propto \pm |k|^z$.
While in two dimensions this leads to the constant conductivity observed in graphene,
in three-dimensional Dirac electron systems with a linearly dispersing cone, this yields
\begin{equation}
\sigma(\omega) = \frac{N}{24}G_0 \frac{\omega}{v_F} = \frac{N e^2}{12 h} \frac{\omega}{v_F} \quad ,
\label{eq:Fermivelocity}
\end{equation}
where $G_0=2e^2/h=7.748\times 10^{-5}~\Omega^{-1}$ is the conductance quantum, $h=\hbar/(2\pi) = 6.626\times 10^{-34}$~Js Planck's constant, and  $N$ is the number of non-degenerate bands.

While for graphene, quasi-crystals or  Cd$_3$As$_2$
the relevant energy range of intraband transitions may extend up to 1~eV,
in the case of \aeti\ the band are much for narrow and tilted. The linear energy dispersion extends only up to $\pm 100$~meV;\cite{Kobayashi09} accordingly the regime of linear optical conductivity will be restricted below approximately 1500~\cm.
Using Eq.~(\ref{eq:Fermivelocity}) with $N=4$ the best fit yields the Fermi velocity $v_F\approx 4.6 \times 10^6$~cm/s, in good agreement with
estimates from magnetotransport measurements\cite{Tajima12b,Monteverde13} and theoretical considerations.\cite{Kobayashi07,Goerbig08}
As sketched in Fig.~\ref{fig:structure}(c), the Dirac cone of \aeti\ is strongly tilted
with a difference in slope by a factor of 10 or more.\cite{Kobayashi09}
In general, optics is not momentum selective and in the case of anisotropic bands
we always probe the lowest velocity $v_F$, independent on polarization.

These considerations pose the question whether our observation of $\sigma(\omega)\propto \omega$ is a fingerprint of the linear band dispersion. By now \aeti\ was always considered a
strictly two-dimensional electron system system, in accordance with the bandstructure\cite{Kino06,Alemany12,Peterseim16a} as demonstrated in Fig.~\ref{fig:Bandstructure}. However, is this still valid for high pressure,
 low temperatures and low energies?

%\clearpage

%\bibliography{alphaET2I3}

\begin{thebibliography}{86}
\bibitem[{rem({\natexlab{a}})}]{remark3}
\bibinfo{note}{The experiments are performed by standard four-probe method with
  the single crystal placed in a copper-beryllium cell filled with Daphne oil,
  similar to the one used for the optical experiments.}
\bibitem[{rem({\natexlab{b}})}]{remark1}
\bibinfo{note}{Most publications only indicate the pressure applied at room
  temperature. Upon cooling, this value changes significantly. Except otherwise
  stated, we consitently quote the actual pressure for any temperature given.
  This implies that the complete suppression of the insulating state might be
  reached at considerably lower pressure than 12-15~kbar, given in previous
  references.}
\bibitem[{rem({\natexlab{c}})}]{remark2}
\bibinfo{note}{As the applied pressure increases, the lattice hardens,
  generally causing an upshift of the vibrational frequencies. This is less
  effective for the molecular vibrations considered here. Nevertheless, our
  estimate of charge disproportionation $\delta\rho$ has to be considered as an
  upper bound.}
\end{thebibliography}

\end{document}